\documentclass[10pt, conference, compsocconf]{IEEEtran}

\usepackage[pdftex]{graphicx}
\usepackage{amsmath}
\usepackage{algorithmic}
\usepackage{array}
\usepackage{subfig}

\usepackage{multirow}
\usepackage{amsmath}

\newtheorem{definition}{Definition}

\newcolumntype{?}[1]{!{\vrule width #1}}

\begin{document}

\title{Discrete Distribution Estimation with Local Differential Privacy: A Comparative Analysis}

\author{\IEEEauthorblockN{Ba Dung Le}
\IEEEauthorblockA{Charles Sturt University, NSW \\
Cyber Security Cooperative Research Centre \\
Australia \\
Email: bdle@csu.edu.au}
\and
\IEEEauthorblockN{Tanveer Zia}
\IEEEauthorblockA{Charles Sturt University, NSW \\
Cyber Security Cooperative Research Centre \\
Australia \\
Email: tzia@csu.edu.au}}

\maketitle

\begin{abstract}

Local differential privacy is a promising privacy-preserving model for statistical aggregation of user data that prevents user privacy leakage from the data aggregator.
This paper focuses on the problem of estimating the distribution of discrete user values with Local differential privacy. 
We review and present a comparative analysis on the performance of the existing discrete distribution estimation algorithms in terms of their accuracy on benchmark datasets.
Our evaluation benchmarks include real-world and synthetic datasets of categorical individual values with the number of individuals from hundreds to millions and the domain size up to a few hundreds of values.
The experimental results show that 
the Basic RAPPOR algorithm generally performs best for the benchmark datasets in the high privacy regime 
while the k-RR algorithm often gives the best estimation in the low privacy regime.
In the medium privacy regime, the performance of the k-RR, the k-subset, and the HR algorithms are fairly competitive with each other and generally better than the performance of the Basic RAPPOR and the CMS algorithms.  

\end{abstract}

\IEEEpeerreviewmaketitle

\section{Introduction}

In the hyper-connected world today, collecting consumer statistics has been a common practice of companies to understand consumers' insights to improve services and products \cite{erlingsson2014rappor,ding2017collecting,adp2017learning}.
The statistical data collection is highly demonstrated in applications of Internet of Things (IoTs) such as Smart Homes, where energy consumption statistics from homeowners can be collected to optimize energy utilization, and Smart Cities, where road transport statistics from motor vehicles can be collected to improve the urban mobility  \cite{miorandi2012internet}.
However, the collection of consumer data requires companies to protect consumer or user privacy to comply with enacted privacy laws and regulation \cite{regulation2016regulation}.

Local differential privacy (LDP) \cite{kasiviswanathan2011can,duchi2013local} is a promising privacy-preserving model for statistical aggregation of user data that prevents user privacy leakage from the data aggregator.
LDP achieves the privacy guarantee by introducing random noise into user data before transmitting them to the data aggregator while maintaining the users' statistics to be accurate. 
Since the data aggregator cannot confidently know the raw user data, the users have plausible deniability and, therefore, their privacy is remained protected to some degree.

Given a mechanism $\mathcal{M}$ as a function of a value $v$ that perturbs $v$ to a value $s$ and returns $s$ as a noisy representation of $v$. 
The formal definition of Local differential privacy is given below.

\begin{definition}[$\epsilon$-Local differential privacy \cite{erlingsson2014rappor}] 
	\label{def_LPD}
	A randomized mechanism $\mathcal{M}$ satisfies $\epsilon$-Local differential privacy if for all pairs of values $v_1$ and $v_2$ given by users, and all set $S$ of the possible outputs of $\mathcal{M}$,
	$$Pr[\mathcal{M}(v_1) \in S] \leq  exp(\epsilon) \times Pr[\mathcal{M}(v_2) \in S]$$
	where 
	the probability space is over the randomness of the mechanism $\mathcal{M}$.
	
\end{definition}

The parameter $\epsilon$, called \textit{privacy lost} or \textit{privacy budget}, takes a real positive value specified by the data aggregator to control the strength of the privacy protection.
The smaller the value of $\epsilon$, the higher the probability that the noisy representations of user values are the same or, in other words, the less likely the data aggregator realizes the true user values. Therefore, a smaller value of $\epsilon$ gives stronger protection of privacy.
However, the smaller the value of $\epsilon$, the worse the accuracy of the aggregated statistics because the noise eliminates useful statistical information about true user data.
Thus, the privacy budget $\epsilon$ needs to be carefully chosen to balance privacy protection and statistical accuracy \cite{hsu2014differential}. 
In practice, $\epsilon$ is generally set to a value from 0.01 to 1 for a high privacy regime and from 1 to 10 for a low privacy regime \cite{erlingsson2014rappor}.

\begin{figure*}[h!]
	\centering
	\includegraphics[scale=0.49]{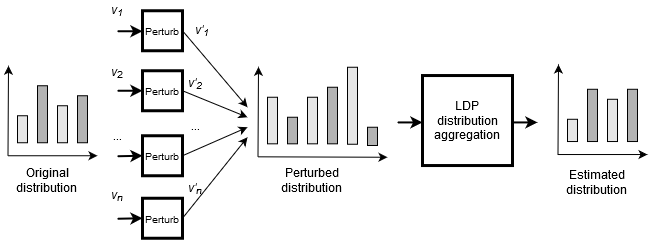}
	\caption{Discrete distribution estimation with Local differential privacy}
	\label{fig:discrete_distribution_estimation}
\end{figure*}

Discrete distribution estimation with LDP is the task of estimating the distribution of discrete user values from privatized data given by LDP mechanisms without assessing the true data \cite{kairouz2016discrete}. 
In this paper, we consider the following scenario of discrete distribution estimation. 
There is a set of $n$ users $U=\{ u_1, u_2, .., u_n \} $ and 
each user $u_i$ has one value taken from a set of $d$ values $V=\{v_1, v_2, .., v_d\}$.
Let $c(v_j)$ being the number of users holding value $v_j$.
The (frequency) distribution of the user values is a vector $f=(f_{v_1}, f_{v_2}, .., f_{v_d})$ with $f_{v_j}$ being $c(v_j)/n$.
Discrete distribution estimation algorithms with LDP perturb user values at user locations and send the privatized data to the data aggregator for distribution aggregation.
Figure \ref{fig:discrete_distribution_estimation} illustrates the steps of discrete distribution estimation algorithms with LDP.

Algorithms for discrete distribution estimation with LDP have been proposed in the literature \cite{kairouz2016discrete,wang2016mutual,erlingsson2014rappor,adp2017learning,acharya2019hadamard}. However, there is a lack of comparison of these algorithms regarding the estimation accuracy and, thus, it is difficult to select a suitable LDP distribution estimation algorithm for a particular dataset.
Some of the existing work \cite{wang2016mutual,acharya2019hadamard} includes a comparison of discrete distribution algorithms but the comparison is solely made based on synthetic datasets that are dissimilar to real-world datasets, such as those are used in our evaluation.

In this paper, we review and present a comparative analysis on the performance of the existing discrete distribution estimation algorithms \cite{kairouz2016discrete,wang2016mutual,erlingsson2014rappor,adp2017learning,acharya2019hadamard} in terms of their accuracy on benchmarks of real-world and synthetic datasets.
The real-world datasets include the Statlog (Australian Credit Approval), the Adult, and the USCensusData1990 datasets that are publicly available at the  UCI Machine Learning Repository \cite{asuncion2007uci}.
These datasets contain categorical individual values with dataset size from hundreds to millions of individuals and domain size up to a few hundred values.
The synthetic datasets are generated to be similar to the real-world benchmark datasets regarding dataset size and domain size.

\section{Algorithms}
\label{frequency_estimation}

\subsection{The Randomized response technique \cite{warner1965randomized}}

The Randomized response (RR) technique is perhaps the earliest proposed algorithm for distribution estimation that guarantees LDP.
This algorithm is originally designed for estimating the distribution of binary values, for example, being "yes" or "no" only.
The two user values can be represented as a single bit with value 1 indicating a user value and value 0 indicating the negation of the user value.

The RR algorithm perturbs a user value $v$ by returning this value with probability $p$ and returning the negation of the value with probability $(1-p)$, for $p \ne 1/2$. 
The perturbation mechanism of the RR algorithm satisfies LDP with the privacy budget $$\epsilon=ln\frac{p}{1-p}.$$

The original distribution of user values is aggregated as follows.
Let $f'_v$ be the proportion of the perturbed values being $v$, the estimated proportion of value $v$ in the true user data is  \cite{warner1965randomized}
$$\hat{f}_v=\frac{f'_v+p-1}{2p-1}.$$

\subsection{The generalized Randomized response algorithm \cite{kairouz2014extremal,kairouz2016discrete}}

The generalized Randomized response algorithm, named k-RR, is a generalization of the RR algorithm to the case that users have more than two values.

Given the privacy budget $\epsilon$, the k-RR algorithm privatizes a user value by sending this value with probability $\frac{e^\epsilon}{e^\epsilon + d -1}$ 
and sending one of the remaining user values with probability $\frac{1}{e^\epsilon + d -1}$ to the data aggregator. 
When $d=2$, the k-RR algorithm is identical to the RR algorithm with the probability $p=\frac{e^\epsilon}{e^\epsilon+1}$.

The k-RR algorithm aggregates the distribution of user values using the maximum likelihood estimator \cite{kairouz2016discrete}. 

\subsection{The k-subset algorithm \cite{wang2016mutual} }
The k-subset algorithm privatizes a user value by sending a set of values sampled from the user value domain to the data curator.
The number of the sampled values is defined by parameter $k$.
For a user value $v$, the privatized set $S_v$ has a probability of $\frac{d e^\epsilon}{ke^\epsilon + d - k}/(^d_k)$ to include $v$ into itself and a probability of $\frac{d }{ke^\epsilon + d - k}/(^d_k)$ to not include $v$.
When $k=1$, the k-subset algorithm is equivalent to the k-RR algorithm.

Given all the privatized sets of user values, the k-subset algorithm aggregates the true distribution of user values as follows.
Let $c'(v)$ be the number of times the value $v$ occurred in a privatized set.
Given $g_k = \frac{ke^\epsilon}{ke^\epsilon+d-k}$ and
$h_k = \frac{ke^\epsilon}{ke^\epsilon+d-k}\frac{k-1}{d-1}+\frac{d-k}{ke^\epsilon+d-k}\frac{k}{d-1}$,
the estimated proportion of the value $v$ in the true user data is \cite{wang2016mutual} 
$$\hat{f}_v = \frac{c'(v)-h_k n}{(g_k - h_k)n}.$$

\subsection{The Basic RAPPOR algorithm \cite{erlingsson2014rappor}}

The Basic RAPPOR algorithm is a modified version of the RAPPOR algorithm \cite{erlingsson2014rappor} that maps a user value to a single bit in a bit sequence, or a one-hot vector.
A one-hot vector representing a value $v$ is a bit sequence with only the bit at the position $v$ set to 1 and the other bits set to 0. 

To privatize a user value, the Basic RAPPOR algorithm applies a mechanism based on the Randomized response technique \cite{warner1965randomized}, simply called randomized response, to perturb every single bit of the represented one-hot vector.
The perturbed bit vector initially has all the bits set to 0. 
For each of the output bits, the bit is set to 1 with probability $q$ if the corresponding bit in the input bit vector is 1 and with probability $p$ if the corresponding bit in the input bit vector is 0.

Given all the privatized bit vectors, the Basic RAPPOR algorithm aggregates the true distribution of user values as follows. 
Let $c'(v)$ be the number of times the bit at the position $v$ in a privatized bit vector is set to 1. 
The proportion of the value $v$ in the true user data is \cite{erlingsson2014rappor}
$$\hat{f}_v = \frac{c'(v) - pn}{(q-p)n}$$

\subsection{The Count Mean Sketch algorithm \cite{adp2017learning}}

The Count Mean Sketch (CMS) algorithm, as similar to the Basic RAPPOR algorithm, 
represents a user value as a one-hot vector and applies randomized response to every bit of the one-hot vector to privatize the user value.
However, the CMS algorithm independently flips each bit of the one-hot vector with probability $\frac{1}{e^{\epsilon/2 + 1}}$ and keep each bit unchanged with the compliment probability. 

The CMS algorithm aggregates the true distribution of user values from perturbed bit vectors as follows.
Let $c'(v)$ be the number of times the bit at the position $v$ in a perturbed bit vector is set to 1. 
Given $c = \frac{e^\epsilon+1}{e^\epsilon-1}$, 
the proportion of the value $v$ in the true user data is \cite{adp2017learning}

$$\hat{f}_v = c'(v)\frac{c+1}{2} + \bigg(c'(v)-n\bigg)\frac{c-1}{2}$$

\subsection{The Hadamard response algorithm \cite{acharya2019hadamard}}
The Hadamard response (HR) algorithm privatizes a user value $v$ in a domain of size $d$ by returning a value $v'$ in the domain of size $d'$ with $d \le d' \le 4d$.
To choose $v'$, suppose $d'$ is a power of two, the HR algorithm first creates a Hadamard matrix $H_{d'} = \{1,-1\}_{d' \times d'}$ as follows:

$H_1 = [+1]$.

$ H_m = 
\begin{bmatrix}
	H_{m/2} & H_{m/2} \\
	H_{m/2} & -H_{m/2}
\end{bmatrix}
$
with $m = 2^j$ for $1\leq j\leq log(d')$.

The HR algorithm then creates a set of values $S_v$ of size $s$ with $s \le d'$.
The set $S_v$ includes all the elements in the row $(v+1)^{th}$ of the Hadamard matrix with row index starting from 0 and at the columns in the matrix with a '+1'.

To privatize $v$, the HR algorithm returns an element of $S_v$ with probability $\frac{e^\epsilon}{se^\epsilon + d' - s}$
and returns an element in the domain of size $d'$ but not in $S_v$ with probability $\frac{1}{se^\epsilon+d'-s}$.
When $d'=d$, $s=1$ and $S_v=\{v\}$, the HR algorithm is equivalent to the k-RR algorithm.

The original distribution of user values is aggregated using the following estimation.
Let $c'(v)$ be the number of the privatized values of $v$ that are in $S_v$. The estimated proportion of $v$ in the true user data is

$$\hat{f}(v) = \frac{2(e^\epsilon+1)}{e^\epsilon-1}\big(\frac{c'(v)}{n}-\frac{1}{2}\big).$$

\section{Experimental results}
In this section, we compare the performance of k-RR, Basic RAPPOR (bRAPPOR), CMS, k-subset, and HR in terms of accuracy on benchmark datasets.
The benchmark datasets should include discrete values specifying user attributes.
The compared algorithms first privatize these values and then estimate the discrete distribution of the original values based on the privatized data, without accessing the original data.

\subsection{Experiment setting}

Previous work having an evaluation of discrete distribution algorithms with LDP test the algorithms on synthetic data only
\cite{kairouz2016discrete,wang2016mutual,acharya2019hadamard} or on real-world datasets but the dataset are not made public \cite{erlingsson2014rappor,adp2017learning}.
We use publicly available real-world datasets as well as the synthetic dataset generated to be similar to these real-world datasets, in terms of dataset size and domain size. 

For real-world datasets, we use three datasets of categorical individual values publicly available at the UC Irvine Machine Learning Repository \cite{asuncion2007uci}. These datasets include the Statlog (Australian Credit Approval), the Adult, and the  USCensusData1990 datasets.
For each dataset, we select only the attributes that have a categorical data type for estimating the distribution of the categorical values in each attribute.
The number of samples in each dataset, the selected attributes, and the value domain size of each attribute are listed in Table \ref{tab:datasets}. 
The attributes of each dataset are listed in the increasing order of the value domain sizes.
For detailed descriptions of these attributes, readers are referred to \cite{asuncion2007uci}.

\begin{table}[h!]
	\centering
	\caption{Real-world datasets with the number of individuals (n), the selected attributes and the value domain size (d).}
	\label{tab:datasets}
	\begin{tabular}{|c|r|c|r|}
		\hline
		Dataset& n &Attribute &d\\ \hline
		
		\multirow{3}{*}{Statlog} & & A4  & 3\\ \cline{3-4}
		& 690 & A6 & 8 \\ \cline{3-4}
		&& A5 &  14\\ \hline

		\multirow{3}{*}{Adult}& &Race & 5 \\ \cline{3-4}
		& 32,561 & Occ & 15 \\ \cline{3-4}
		&&Country & 42 \\ \hline

		\multirow{3}{*}{USCensus1990}& &Military & 5 \\ \cline{3-4}
		& 2,458,285 & Rvetserv & 12 \\ \cline{3-4}
		&&Race & 63 \\ \cline{3-4}
		&& PoB & 283 \\ \hline
		
	\end{tabular}
	
\end{table}

We generate synthetic datasets of categorical values following the Geometric distribution.
For the synthetic datasets with the number of samples up to 20,000 (small datasets), 
the generated values are taken from the two categorical domains of 20 and 100 values.
For the synthetic datasets with the number of samples from 20,000 to 400,000 (large datasets), 
the generated values are taken from the two categorical domains of 100 and 500 values.

\begin{table*}[t!]
	\centering
	\caption{Performance (MAE) of the compared algorithms on the real world datasets for $\epsilon = 0.5$.}
	\label{tab:high_privacy}
	\begin{tabular}{|c|c?{0.4mm}c|c?{0.4mm}c|c?{0.4mm}c|c?{0.4mm}c|c?{0.4mm}c|c?{0.4mm}}
		\hline
		\multirow{2}{*}{Dataset}& \multirow{2}{*}{Attribute}& \multicolumn{2}{c?{0.4mm}}{\textbf{k-RR}} & \multicolumn{2}{c?{0.4mm}}{\textbf{bRAPPOR}}  & \multicolumn{2}{c?{0.4mm}}{\textbf{CMS}} & \multicolumn{2}{c?{0.4mm}}{\textbf{k-subset}} & \multicolumn{2}{c?{0.4mm}}{\textbf{HR}}\\ \cline{3-12}
		&  & Mean & Std.& Mean & Std. & Mean & Std. & Mean & Std. & Mean & Std.\\ \hline
		
		\multirow{2}{*}{Statlog} & A4 &0.067 & 0.036 & \textbf{0.05} & 0.024 & 0.109 & 0.054 & 0.063 & 0.037 & 0.078 & 0.048 \\ \cline{2-12}
		& A6 &0.083 & 0.03 & \textbf{0.043} & 0.014 & 0.089 & 0.03 & 0.082 & 0.022 & 0.071 & 0.023 \\ \cline{2-12}
		& A5  &0.083 & 0.013 & \textbf{0.05} & 0.01 & 0.086 & 0.021 & 0.085 & 0.014 & 0.061 & 0.011 \\ \hline

		\multirow{2}{*}{Adult}&Race   &0.012 & 0.005 & \textbf{0.008} & 0.002 & 0.015 & 0.006 & 0.013 & 0.004 & 0.014 & 0.005 \\ \cline{2-12}
		& Occ  &0.024 & 0.005 & \textbf{0.009} & 0.002 & 0.016 & 0.003 & 0.015 & 0.003 & 0.015 & 0.003 \\ \cline{2-12}
		&Country   &0.008 & 0.002 & \textbf{0.005} & 0.001 & 0.01 & 0.002 & 0.01 & 0.001 & 0.013 & 0.002 \\ \hline

		\multirow{2}{*}{USCensus1990}&Military   &0.002 & 0.001 & \textbf{0.001} & 0.0 & 0.002 & 0.001 & 0.002 & 0.001 & 0.002 & 0.001 \\ \cline{2-12}
		& Rvetserv   &0.003 & 0.001 & \textbf{0.001} & 0.0 & 0.002 & 0.0 & 0.002 & 0.0 & 0.002 & 0.001 \\ \cline{2-12}
		&Race  &\textbf{0.001} & 0.0 & 0.016 & 0.0 & 0.023 & 0.0 & \textbf{0.001} & 0.0 & 0.002 & 0.0 \\ \cline{2-12}
		& PoB &0.004 & 0.0 & 0.004 & 0.0 & 0.005 & 0.0 & \textbf{0.001} & 0.0 & \textbf{0.001} & 0.0 \\ \hline
		
	\end{tabular}
	
\end{table*}

\begin{table*}[t!]
	\centering
	\caption{Performance (MAE) of the compared algorithms on the real world datasets for $\epsilon = 1$.}
	\label{tab:medium_privacy}
	\begin{tabular}{|c|c?{0.4mm}c|c?{0.4mm}c|c?{0.4mm}c|c?{0.4mm}c|c?{0.4mm}c|c?{0.4mm}}
		\hline
		\multirow{2}{*}{Dataset}& \multirow{2}{*}{Attribute}& \multicolumn{2}{c?{0.4mm}}{\textbf{k-RR}} & \multicolumn{2}{c?{0.4mm}}{\textbf{bRAPPOR}}  & \multicolumn{2}{c?{0.4mm}}{\textbf{CMS}} & \multicolumn{2}{c?{0.4mm}}{\textbf{k-subset}} & \multicolumn{2}{c?{0.4mm}}{\textbf{HR}}\\ \cline{3-12}
		&  & Mean & Std.& Mean & Std. & Mean & Std. & Mean & Std. & Mean & Std.\\ \hline		
		
		\multirow{2}{*}{Statlog} & A4   &\textbf{0.033} & 0.02 & 0.05 & 0.023 & 0.044 & 0.025 & 0.034 & 0.02 & 0.106 & 0.04 \\ \cline{2-12}
		& A6 &\textbf{0.041} & 0.013 & 0.044 & 0.013 & 0.043 & 0.014 & 0.044 & 0.013 & 0.048 & 0.015 \\ \cline{2-12}
		& A5   &0.054 & 0.011 & 0.049 & 0.011 & 0.049 & 0.01 & 0.051 & 0.009 & \textbf{0.046} & 0.008 \\ \hline		
		
		\multirow{2}{*}{Adult}&Race  &	\textbf{0.006} & 0.002 & 0.008 & 0.003 & 0.008 & 0.003 & 0.007 & 0.002 & 0.008 & 0.003 \\ \cline{2-12}
		& Occ   &0.01 & 0.002 & \textbf{0.008} & 0.002 & \textbf{0.008} & 0.002 & \textbf{0.008} & 0.002 & 0.016 & 0.003 \\ \cline{2-12}
		&Country   & \textbf{0.005} & 0.001 & \textbf{0.005} & 0.001 & \textbf{0.005} & 0.001 & 0.006 & 0.001 & 0.008 & 0.001 \\ \hline			    
		
		\multirow{2}{*}{USCensus1990}&Military   &\textbf{0.001} & 0.0 & \textbf{0.001} & 0.0 & \textbf{0.001} & 0.0 & \textbf{0.001} & 0.0 & \textbf{0.001} & 0.0 \\ \cline{2-12}
		& Rvetserv   &\textbf{0.001} & 0.0 & \textbf{0.001} & 0.0 & \textbf{0.001} & 0.0 & \textbf{0.001} & 0.0 & \textbf{0.001} & 0.0 \\ \cline{2-12}
		&Race   &\textbf{0.001} & 0.0 & 0.016 & 0.0 & 0.021 & 0.0 & \textbf{0.001} & 0.0 & 0.01 & 0.0 \\ \cline{2-12}
		& PoB   & 0.002 & 0.0 & 0.004 & 0.0 & 0.005 & 0.0 & \textbf{0.001} & 0.0 & \textbf{0.001} & 0.0 \\ \hline
		
	\end{tabular}
	
\end{table*}

\begin{table*}[t!]
	\centering
	\caption{Performance (MAE) of the compared algorithms on the real world datasets for $\epsilon = 2$.}
	\label{tab:low_privacy}
	\begin{tabular}{|c|c?{0.4mm}c|c?{0.4mm}c|c?{0.4mm}c|c?{0.4mm}c|c?{0.4mm}c|c?{0.4mm}}
		\hline
		\multirow{2}{*}{Dataset}& \multirow{2}{*}{Attribute}& \multicolumn{2}{c?{0.4mm}}{\textbf{k-RR}} & \multicolumn{2}{c?{0.4mm}}{\textbf{bRAPPOR}}  & \multicolumn{2}{c?{0.4mm}}{\textbf{CMS}} & \multicolumn{2}{c?{0.4mm}}{\textbf{k-subset}} & \multicolumn{2}{c?{0.4mm}}{\textbf{HR}}\\ \cline{3-12}
		&  & Mean & Std.& Mean & Std. & Mean & Std. & Mean & Std. & Mean & Std.\\ \hline
		
		\multirow{2}{*}{Statlog} & A4   &\textbf{0.013} & 0.008 & 0.047 & 0.024 & 0.024 & 0.013 & \textbf{0.013} & 0.008 & 0.031 & 0.013 \\ \cline{2-12}
		& A6  & \textbf{0.018} & 0.005 & 0.047 & 0.014 & 0.025 & 0.006 & 0.027 & 0.008 & 0.029 & 0.008 \\ \cline{2-12}
		& A5   & \textbf{0.022} & 0.004 & 0.051 & 0.01 & 0.027 & 0.006 & 0.031 & 0.006 & 0.032 & 0.007 \\ \hline
		
		\multirow{2}{*}{Adult}&Race  & \textbf{0.003} & 0.001 & 0.008 & 0.003 & 0.004 & 0.002 & 0.004 & 0.001 & 0.055 & 0.002 \\ \cline{2-12}
		& Occ   & \textbf{0.003} & 0.001 & 0.009 & 0.002 & 0.004 & 0.001 & 0.005 & 0.001 & 0.006 & 0.001 \\ \cline{2-12}
		&Country   &\textbf{0.003} & 0.0 & 0.005 & 0.001 & \textbf{0.003} & 0.0 & 0.004 & 0.0 & 0.007 & 0.001 \\ \hline		  
		
		\multirow{2}{*}{USCensus1990}&Military  &  \textbf{0.0} & 0.0 & 0.001 & 0.0 & 0.001 & 0.0 & 0.001 & 0.0 & 0.039 & 0.0 \\ \cline{2-12}
		& Rvetserv   &\textbf{0.0} & 0.0 & 0.001 & 0.0 & 0.001 & 0.0 & 0.001 & 0.0 & 0.0 & 0.0 \\ \cline{2-12}
		&Race   & \textbf{0.0} & 0.0 & 0.016 & 0.0 & 0.018 & 0.0 & 0.0 & 0.0 & 0.003 & 0.0 \\ \cline{2-12}
		& PoB   &    0.001 & 0.0 & 0.004 & 0.0 & 0.004 & 0.0 & \textbf{0.0} & 0.0 & 0.001 & 0.0 \\ \hline
		
	\end{tabular}
	
\end{table*}

We implement the k\_RR, the Basic RAPPOR, the CMS, and the k-subset algorithms following \cite{kairouz2016discrete,wang2016mutual,erlingsson2014rappor,adp2017learning}. The implementation of the Hadamard algorithm is publicly provided as listed in \cite{acharya2019hadamard}.
The algorithms are executed with default values of parameters.
The performance of the algorithms are compared in three privacy regimes with $\epsilon$ = 0.5 for high privacy regime, $\epsilon$ = 1 for medium privacy regime, and $\epsilon$ = 2 for low privacy regime. 

The accuracy of the distribution estimation is measured by Mean Absolute Error (MAE) which is defined as
$$MAE(\hat{f},f) =\frac{1}{d} \sum^d_{i=1}|\hat{f}_{v_i} - f_{v_i}|$$ 
where
$\hat{f}_{v_i}$ and $f_{v_i}$ are the estimated and the true proportions of user value $v_i$ respectively.
The MAE value is the average of the absolute differences between all the elements of the estimated distribution and the true distribution of user values.
The smaller the MAE value, the more accurate the distribution estimation.

\subsection{Results}
Table \ref{tab:high_privacy}, Table \ref{tab:medium_privacy} and Table \ref{tab:low_privacy} show the accuracy of the evaluated algorithms on the real world datasets for the three privacy regimes.
The result for the Statlog and the Adult datasets is the average of the results from a hundred runs.
The result of the USCensus dataset is the average of the results from 10 runs.
The column Mean lists the average of the MAE values obtained from all the runs of the algorithm and the column Std. lists the standard deviation of the MAE values.
The higher the value of Std., the more fluctuated the estimation accuracy.
The best performing results (or the lowest MAE values) are highlighted in bold.

As can be seen from the tables, the Basic RAPPOR algorithm performs best for most of the datasets in the high privacy regime ($\epsilon= 0.5$).
In the medium privacy regime ($\epsilon= 1$), k-RR performs better than the other algorithms for the datasets with a small number of samples 
while on the datasets with a large number of samples, k-subset and HR are fairly competitive with k-RR.
In the low privacy regime ($\epsilon= 2$), the k-RR algorithm generally performs better than the other algorithms.

\begin{figure*}[t!]
	\subfloat[][$\epsilon$=0.5]{
		\includegraphics[scale=0.42]{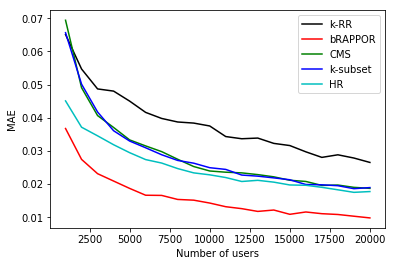}
	} 	
	\subfloat[][$\epsilon$=1]{
		\includegraphics[scale=0.42]{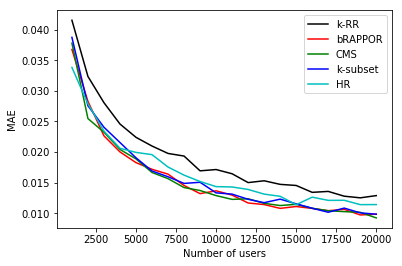}
	} 	
	\subfloat[][$\epsilon$=2]{
		\includegraphics[scale=0.42]{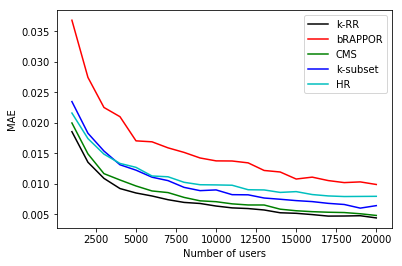}
	} 	
	\caption{Performance (MAE) of the compared algorithms on the small synthetic datasets ($n \leq 20,000$) with d=20.}
	\label{fig:synthetic_smalld20}
\end{figure*}

\begin{figure*}[t!]
    \subfloat[][$\epsilon$=0.5]{
		\includegraphics[scale=0.42]{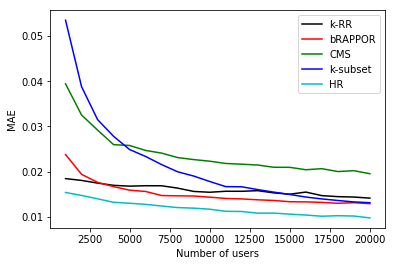}	
	}
	\subfloat[][$\epsilon$=1]{
		\includegraphics[scale=0.42]{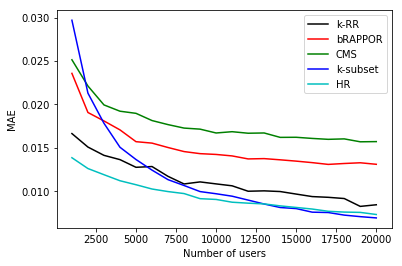}	
	}
	\subfloat[][$\epsilon$=2]{
		\includegraphics[scale=0.42]{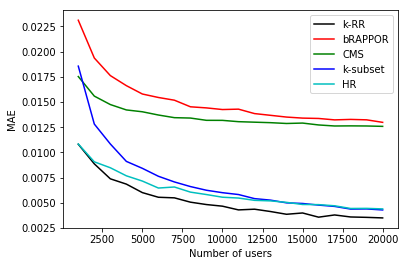}	
	}
	\caption{Performance (MAE) of the compared algorithms on the small synthetic datasets ($n \leq 20,000$) with d=100.}
	\label{fig:synthetic_smalld100}
\end{figure*}

\begin{figure*}[t!]
	\subfloat[][$\epsilon$=0.5]{
		\includegraphics[scale=0.42]{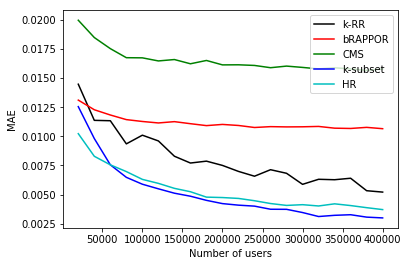}
	} 	
	\subfloat[][$\epsilon$=1]{
		\includegraphics[scale=0.42]{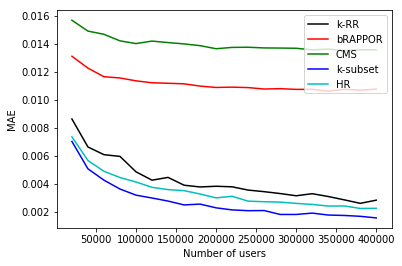}
	} 	
	\subfloat[][$\epsilon$=2]{
		\includegraphics[scale=0.42]{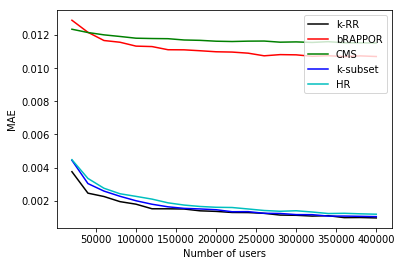}
	} 	
	\caption{Performance (MAE) of the compared algorithms on the large synthetic datasets ($20,000 < n \leq 400,000$) with d=100.}
	\label{fig:synthetic_larged100}
\end{figure*}

\begin{figure*}[t!]
    \subfloat[][$\epsilon$=0.5]{
		\includegraphics[scale=0.42]{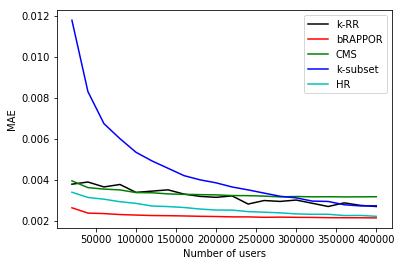}	
	}
	\subfloat[][$\epsilon$=1]{
		\includegraphics[scale=0.42]{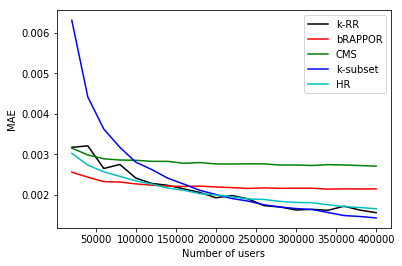}	
	}
	\subfloat[][$\epsilon$=2]{
		\includegraphics[scale=0.42]{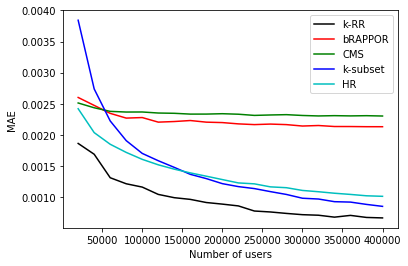}	
	}
	\caption{Performance (MAE) of the compared algorithms on the large synthetic datasets ($20,000 < n \leq 400,000$) with d=500.}
	\label{fig:synthetic_larged500}
\end{figure*}

Figure \ref{fig:synthetic_smalld20}, Figure \ref{fig:synthetic_smalld100}, 
Figure \ref{fig:synthetic_larged100} and Figure \ref{fig:synthetic_larged500}
illustrate the performance of the compared algorithms on the synthetic datasets.
According the figures, in the high privacy regime ($\epsilon=0.5$), 
Basic RAPPOR generally outperforms the other algorithms for the small datasets with small value domain size (Figure \ref{fig:synthetic_smalld20}a) 
and for the large datasets with large value domain size (Figure \ref{fig:synthetic_larged500}a).
HR is better than the other algorithms for the small datasets with large value domain size (Figure \ref{fig:synthetic_smalld100}a)
while k-subset is the best performing algorithm for the large datasets with small value domain size (Figure \ref{fig:synthetic_larged100}a).

In the medium privacy regime ($\epsilon=1$), 
k-RR, k-subset, and HR are competitive with each other and generally give the best estimation.
In the low privacy regime ($\epsilon=2$), 
k-RR generally has the highest accuracy but the performance of k-subset and HR are often comparable with the performance of k-RR.

The experimental results indicate that the relative accuracy of the compared algorithms are dependent on the characteristics of the evaluation datasets, such as dataset size and value domain size, as well as the privacy protection level.
The dependency of the algorithm performance on dataset characteristics has also been previously observed in \cite{acharya2019hadamard} for a set of synthetic datasets. 
However, our results further confirm the finding for real-world datasets and a set of synthetic datasets with similar characteristics to these real-world data.

The performance dependency of the compared algorithms on dataset characteristics is likely due to their approach to privatizing user values.
While k-RR and k-subset privatize user values by directly perturbing the domain values only, 
HR perturbs user values using the domain values and a pre-defined number of the values outside the user value domain.
Basic RAPPOR and CMS privatize user values using one-hot vector representations representing a large number of the values outside the user value domain.
The perturbation steps of the algorithms differently introduce noise into their privatized data depending on value domain size, dataset size, and privacy protection level.
The estimation accuracy of the algorithms are, therefore, differently sensitive to these parameters.

\section{Conclusions}

We have presented a comparative analysis on the performance of discrete distribution estimation algorithms with Local differential privacy on real-world and synthetic datasets.
The real-world datasets include three datasets of categorical individual values publicly available at the UC Irvine Machine Learning Repository \cite{asuncion2007uci}: the Statlog (Australian Credit Approval), the Adult, and the USCensusData1990 datasets.
The synthetic datasets are generated to be similar to the real-world datasets regarding dataset size and domain size.
We conclude that 
the Basic RAPPOR algorithm generally performs best, in terms of estimation accuracy, for the evaluation datasets in the high privacy regime.
In the medium privacy regime, k-RR, k-subset, and HR give fairly comparable results but and generally are better than Basic RAPPOR and CMS. 
In the low privacy regime, the k-RR algorithm often gives the best estimation.

Our empirical evaluation is based on the execution of the algorithms with default values of parameters. 
To further understand the relative performance of the algorithms, it would be necessary to compare the algorithms for different parameter settings. 
For example, the accuracy of the Basic RAPPOR algorithm would be improved by adopting the optimal choice of parameter values as discussed in \cite{wang2017locally}.
In the limited scope of this paper, we have only focused on the estimation accuracy of the algorithms.  
Further work will be required to evaluate the algorithms in other aspects such as computational complexity and communication cost.

\section*{Acknowledgment}

The work has been supported by the Cyber Security Research Centre Limited whose activities are partially funded by the Australian Government’s Cooperative Research Centres Programme.
We would like to thank anonymous reviewers for their constructive
and helpful comments on an earlier version of the manuscript.

\bibliographystyle{plain}
\bibliography{Local_differential_privacy}

\end{document}